\documentclass[12pt]{iopart}
\usepackage{amssymb}

\usepackage{graphicx}
\usepackage{color}
\begin{document}

\rapid{Weakly Bound Cluster States of Efimov Character}
\author{Javier von Stecher}
\address{ JILA, University of Colorado and National Institute of Standards and Technology, Boulder, CO 80309-0440}

\begin{abstract}
We study the behavior of weakly bound clusters and their relation to the well-known three-body Efimov states.
We adopt a model to describe universal behavior of strongly interacting bosonic systems, and we
test its validity by reproducing predictions of three- and four-body universal states.
Then, we extend our study to larger systems and identify a series of universal cluster states that can be qualitatively interpreted as adding one particle at a time to an Efimov trimer. The properties of these cluster states and their experimental signatures are discussed.

\end{abstract}

\maketitle

The universal behavior of
quantum systems has attracted much attention from the atomic, condensed-matter, and nuclear physics communities.
Theoretical and experimental studies
strongly suggest that the ultracold behavior of two-component Fermi gases is characterized by a single interaction parameter: the scattering length $a$ (cf. Ref.~\cite{giorgini2008tua} and references therein).
In contrast, there is no comprehensive understanding of universality in bosonic systems with large $a$.  For weakly interacting dilute systems, there exists a gaslike state in which the interactions depend on $a$. However, this state becomes unstable against collapse as $a$ increases~\cite{donley2001dynamics}.
 At the three-body level ($N$=3), Efimov physics~\cite{efim70} leads to a different concept of universality. In this case, two parameters are necessary to describe the low-energy behavior: $a$ and the three-body parameter, $\kappa_0$, that is related to the energy of an Efimov state at unitarity ($|a|=\infty$). The experimental observation of Efimov phenomena in ultracold gases~\cite{kraemer2006eeq} has reinvigorated the interest in universal few-body physics.

The extension of Efimov physics to larger systems has been an issue of debate. Initial studies~\cite{FJ}, based on restricting approximations, proposed the existence of an ``$N$-body Efimov effect'' for systems with $N>3$. This prediction does not agree with a full quantum-mechanical treatment of the four-boson problem~\cite{von2009signatures}. Recent four-body studies~\cite{von2009signatures,hammer2007upf,Schmidt09} predict that the system is universal, i.e., no new parameter is needed, and the low-energy behavior of the system is only described by $a$ and $\kappa_0$. The experimental observation of resonant four-body features~\cite{ferlaino2009evidence}, in agreement with theoretical predictions of Ref.~\cite{von2009signatures}, strengthens our confidence in the understanding of universal four-body phenomena.
Another series of studies focused on the existence of weakly bound cluster states. These studies considered such systems as $^4$He and tritium or others that interact through realistic van der Waals or other model potentials~\cite{blume2000mch,blume2002formation,lewerenz1997structure,hanna2006energetics}.
Although there are clear differences between these studies, their predictions are qualitatively similar, possibly indicating an underlying universal behavior. Furthermore, the studies of Ref.~\cite{hanna2006energetics} suggest that cluster states only depend on two- and three-body physics. This opens up the possibility of predicting universal properties of cluster states only in terms of $a$ and $\kappa_0$~\cite{hammer2007upf}.

Here, we propose a method to explore the
 properties of bosonic systems with large $a$. We adopt a model Hamiltonian that combines two- and three-body interactions and allows the independent control of $a$ and $\kappa_0$. Our model Hamiltonian is based on Efimov solutions of the three-body problem and designed to significantly reduce the nonuniversal corrections of the lowest energy states. This property makes it particularly suitable for studying larger systems for which only the lowest states are computationally accessible.
  Using numerical techniques, such as diffusion Monte Carlo (DMC) and correlated-Gaussian (CG) basis-set expansion, we study the properties of  clusters.
   First, we show that the model Hamiltonian reproduces the universal behavior of three- and four-body systems.
 Second, we extend our calculations to systems with $N\le13$ particles, and we obtain results consistent with the premise of universality.
We conclude that for each Efimov state, there is at least one $N$-body cluster state.
 These  $N$-body cluster states are only controlled by two- and three-body physics and inherit many of the properties of the universal three- and four-body states.
The cluster states are formed by long-range binding and fall into the category of quantum halos~\cite{riisager2000quantum} since most of their probability is in the classically forbidden region.
 For large $a$, the energies of different cluster states are linearly related, forming Tjon lines~\cite{tjon}. Also, under certain conditions, the $N$-body cluster is bound, while all smaller clusters are unbound. This behavior manifests the Borromean nature of the states~\cite{blume2002formation}.
 Our predictions describe states of physical systems that fulfill the conditions of universality: large $a$  ($|a|$ larger than any other length scale of the interactions) and long-range binding. In experimental studies, these conditions might not be fully satisfied by the lowest Efimov family, but can be quantitatively fulfilled by excited families whose cluster states are resonances and, therefore, have a finite width.

{\it Model Hamiltonian.}---
To construct a model Hamiltonian that captures the essence of Efimov physics, we consider the analytical solutions of a three-body system with zero-range interactions in the hyperspherical framework~\cite{braaten2006ufb}. In this representation,  the description of the system reduces to a one-dimensional Schr\"odinger equation in the hyperradius $R=\sqrt{(r_{12}^2+r_{13}^2+r_{23}^2)/3}$. At unitarity ($|a|=\infty$), the description of Efimov physics is governed by an effective
 potential $U(R)=-(s_0^2+1/4)\hbar^2/(2mR^2)$, where $m$ is the mass of the particles and $s_0\approx1.00624$. The divergence of this effective potential at $R\rightarrow 0$ leads to the unphysical Thomas collapse~\cite{thomas35}, i.e., an infinite number of states with energies approaching $-\infty$. Thomas collapse is an artifact of the zero-range interaction being modified when potentials with finite range $r_0$ are considered. Finite-range corrections alter the behavior of $U(R)$ in regions $R\lesssim r_0$, producing deviations from its universal $1/R^2$ behavior and introducing a natural regularization. These finite-range effects strongly affect the lowest energy states, introducing important nonuniversal corrections.

\begin{figure}[h]
\begin{center}
\includegraphics[scale=0.6,angle=0]{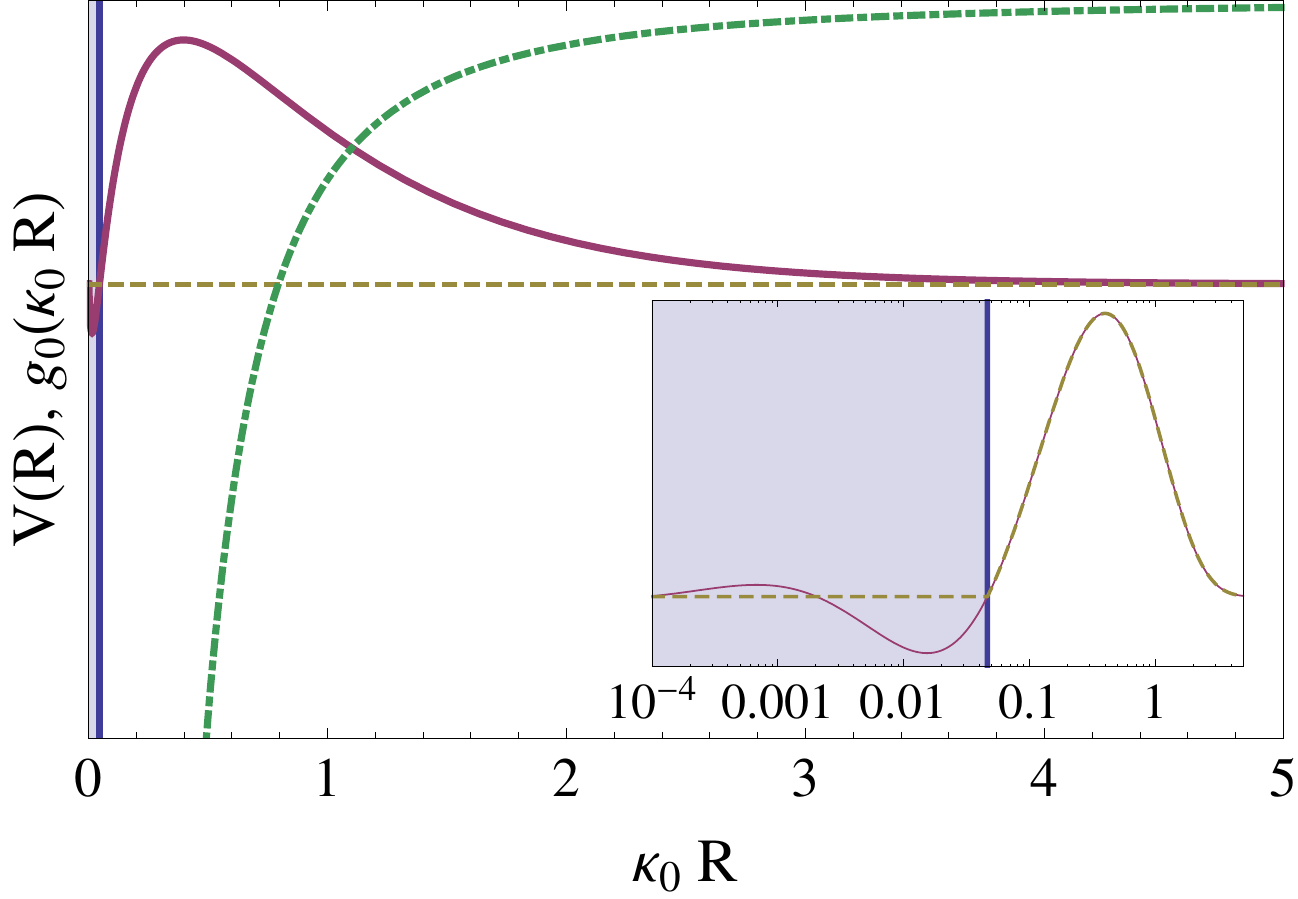}
\end{center}
\caption{(Color online) Description of the model potential in the hyperspherical framework. The vertical line and the shaded region at $\kappa_0 R\ll 1$ represent the hard-wall three-body potential (also included in the inset). The dashed line corresponds to the binding energy, and the solid red line to the wave function $g_0$. The dash-dotted curve corresponds to the potential $U(R)$. Inset: Comparison of  $g_0$ (solid curve) and $\overline{g_0}$ (dashed curve) in logarithmic scale.
} \label{modelFig}
\end{figure}
To obtain universal behavior even for the lowest energy states, we adopt a regularization scheme that consists of introducing a repulsive three-body interaction in the hyperradius. The simplest three-body interaction is a hard wall of size $R_c$, i.e., $V_3=\infty$ for $R<R_c$ and 0 otherwise. Thus, our hyperradial potential at unitarity is $U^M(R)=-(s_0^2+1/4)\hbar^2/(2mR^2)$ for $R>R_c$, and  $U^M(R)=\infty$ otherwise. Note that $U^M(R)$ agrees with $U(R)$ for all $R>R_c$. Thus, the analytical hyperradial solution from Ref.~\cite{braaten2006ufb}, $g_0(\kappa R)=\sqrt{\kappa R}\mbox{K}_{i s_0} (\sqrt{2} \kappa R)$, is valid for our model Hamiltonian for all $R>R_c$. Here, $\mbox{K}_{i s_0}$ is the modified Bessel function of the second kind with imaginary index $i s_0$,  and $\kappa$ is the momentum associated with the energy of the state by $E^U_{3b}=-\kappa^2\hbar^2/m$ (the superscript U refers to unitarity).
 The hard-wall potential 
  imposes the boundary condition $g_0(\kappa_n R_c)=0$ that determines the discrete spectrum $E^U_{3b,n}=-\kappa_n^2\hbar^2/m$. The hyperradial wave function of our model Hamiltonian $\overline{g_0}$ is then $g_0(\kappa_n R)$ for all $R> R_c$ and 0 otherwise.

Figure~\ref{modelFig} summarizes the basic idea of this regularization scheme. Note that only a very small fraction of the wave function $g_0$ is in the region $R<R_c$ (see inset of Fig.~\ref{modelFig}), suggesting that even the ground state of our model Hamiltonian is, to a large extent, universal.
  For example, the lowest energies follow closely the Efimov-scaling relation $E^{U}_{3b,n+1}=e^{-2\pi/s0}E^{U}_{3b,n}$.
The  lowest momenta, $\kappa_0 \approx 0.04622/R_c$ and $\kappa_1 \approx0.0020359/R_c$, imply that the  Efimov-scaling relation is obeyed even for the lowest two states within 0.1\%.
The long-range binding is another essential signature of universality.
The typical interparticle distances are much larger than the interaction range $R_c$, which implies that the properties of the state are insensitive to nonuniversal short-range physics.
These properties suggest that even the lowest state can be considered universal and of Efimov character.

To extend the description of universal physics to larger systems, we adopt the simplest Hamiltonian that leads to the N-body ground state with long-range binding:
\begin{eqnarray}
\label{eq_ham} H = \sum_{i} \frac{-\hbar^2}{2m}\nabla_i^2
+\sum_{i<j} V_2(r_{ij})+\sum_{i<j<k} V_3(R_{ijk}).
\label{Ham}
\end{eqnarray}
Here $r_{ij}$ is the interparticle distance between particles $i$ and $j$, and $R_{ijk}=\sqrt{(r_{ij}^2+r_{ik}^2+r_{jk}^2)/3}$ is the three-body hyperradius.
The three-body force prevents the formation of tightly bound clusters and guarantees long-range binding.
This premise is confirmed {\it a posteriori} when we analyze the pair-correlation function of the cluster states and suggests that $N$-body forces (with $N>3$) are not essential for the description of universal physics.

In Eq.~\ref{Ham}, the ideal combination of interactions would be a zero-range pseudopotential,  $V_2(\mathbf{r})=4\pi a\hbar^2/m \,\delta(\mathbf{r}) \frac{\partial}{\partial r} r$, and a hard-wall potential of size $R_c$ for $V_3$.
Numerically, however, these interaction potentials can be hard to deal with; we need to replace them with slightly different potentials that are better suited to the techniques we apply.
For example, the zero-range pseudopotential has to be replaced by a finite-range potential. This finite-range potential leads to the same two-body physics as the zero-range pseudopotential as long as $|a|\gg r_0$. For CG calculations, we use $V_2(r)=-V^{CG}_{0} e^{-r^2/2r^2_0}$;  for  DMC calculations, we use a
square-well potential: $V_2(r)=-V_0$ for $r<r_0$, and 0 otherwise.
For DMC calculations, we can use the hard-wall three-body potential
 as the three-body interaction, but for CG calculations we need to replace it with
 $V_3(R)=V' e^{-R^2/2R^2_c}$~\cite{von2009signatures}. For this model, universality is expected in the regime $|a|\gg R_c, r_0$. To reduce finite-range corrections, we impose the condition that $r_0\ll R_c$.

{\it Numerical Calculations.}---
To study a system with $N\le 6$, we use a CG basis set expansion~\cite{suzuki1998sva,stech07}. In our implementation, the eigenstates of a system are expanded in a set of CG basis functions in which the center of mass has been removed, and the relative angular momentum is zero.  Each basis function is a symmetrized product of Gaussian functions, each of which depends on one of the $N(N-1)/2$ interparticle distances. The parameters that characterize the Gaussian functions are selected and optimized using a stochastical variational method~\cite{suzuki1998sva}. This optimization allows us to reach convergence of the lowest energy states with a few hundred (thousands) basis functions for $N=3$ ($N=4$, 5).

To extend our calculations to larger systems, we implement a DMC algorithm~\cite{hamm94}. To within statistical uncertainties, the
DMC algorithm provides the exact ground-state energy. We introduce importance sampling through a trial wave function $\Psi_T$ to reduce the statistical uncertainty. The trial wave function is first
 optimized using variational Monte Carlo (VMC) methods.
To describe the weakly bound cluster states,
 we propose the trial wave function:
\begin{equation}
\Psi_T(\mathbf{r}_1,...,\mathbf{r}_n)=\Phi(R_T)\prod_{i<j<k} g(R_{i,j,k}) \prod_{i<j} f(r_{i,j}),
\label{SchrEq}
\end{equation}
where $f$ and $g$ are the two- and three-body correlations, $\Phi$ is a hyperradial correlation, $R_T$ is the $N$-body hyperradius given by $R_T^2=\sum (\mathbf{r}_i-\mathbf{R}_{CM})^2$, and $\mathbf{R}_{CM}=\sum \mathbf{r}_i/N$ is the center-of-mass coordinate. The two-body correlation is the zero-energy scattering solution of two particles interacting through $V_2$, i.e., $f(r)=(r_0-a)\sin(K_0 r)/[r\sin(K_0 r_0)]$ for $r<r_0$, and $f(r)=1-a/r$ otherwise. Here $K_0^2=mV_0/\hbar^2$.
The three-body correlation is $g(R)=0$ for $R<R_c$, and $g(R)=(R-R_c) R^b \exp(-c R^2)$ otherwise. Here $b$ and $c$ are variational parameters.
 The hyperradial correlation is $\Phi(R)=R^{b'} \exp(-c' R^2)$ where $b'$ and $c'$ are variational parameters.

\begin{figure}[h]
\begin{center}
\includegraphics[scale=0.6,angle=0]{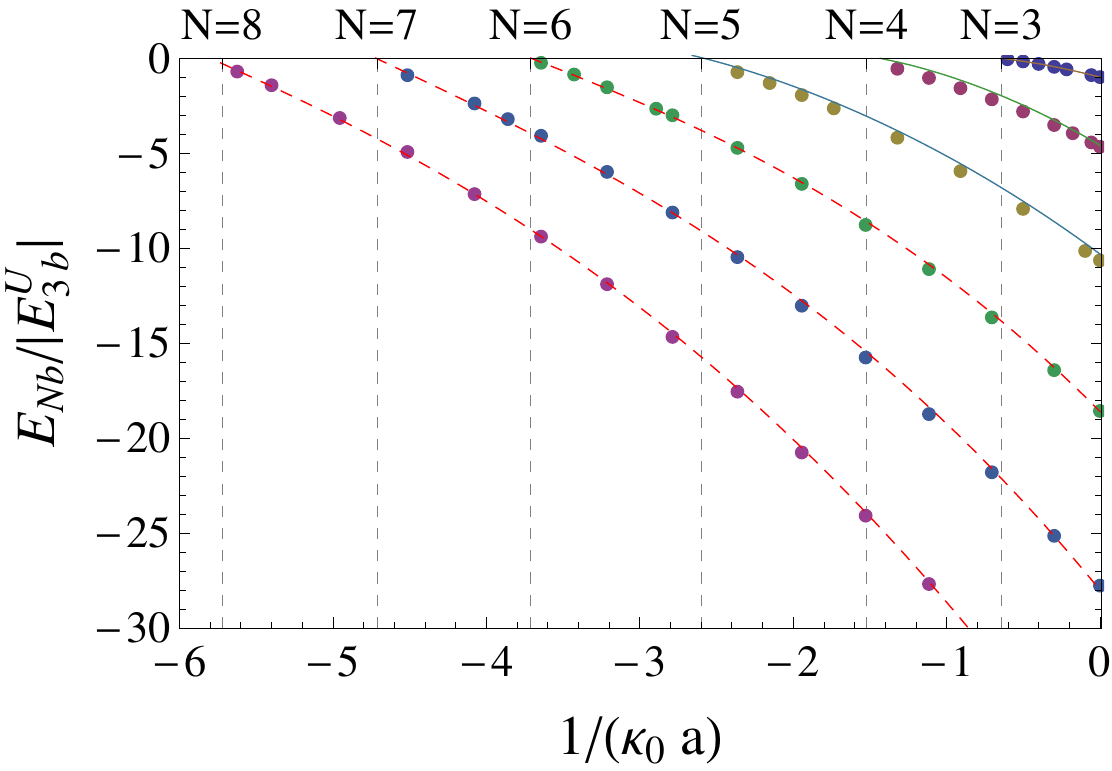}
\end{center}
\caption{(Color online) Bosonic cluster energies in the negative scattering-length region. Symbols correspond to the DMC predictions for $R_c/r_0=4$. The numbers at the top identify the cluster size, and the vertical dashed lines correspond to $1/(\kappa a^*_{Nb})$. For $N=3$ and 4, the solid curves correspond to predictions of Ref.~\cite{braaten2006ufb} and Ref.~\cite{von2009signatures}, respectively. For $N=5$, the solid curve corresponds to CG predictions using our model Hamiltonian. Dashed curves are fits to an analytic simple form (see text). } \label{EnNeg}
\end{figure}

{\it Universal Cluster States.}---We extend our studies to larger systems by solving the many-body Schr\"odinger equation of our model Hamiltonian using CG and DMC methods. Figure~\ref{EnNeg} presents the ground-state energies $E_{Nb,0}\equiv E_{Nb}$ of the bosonic cluster state.
Considering that these universal clusters are only controlled by two- and three-body physics, their energies should obey
 $E_{Nb}= \kappa_0^2\hbar^2/m\,\epsilon_N [1/(\kappa_0 a)]=|E_{3b}^U| \epsilon_N [1/(\kappa_0 a)]$, where the $\epsilon_N$ is a universal function that only depends on $N$. Thus, the numerical results as presented in Fig.~\ref{EnNeg} represent the universal relationship of the bosonic cluster energies given by $\epsilon_N$.
 The symbols represent the Monte Carlo results, while the solid curves correspond to other predictions. For $N=3$, the curve corresponds to the semi analytical prediction of Ref.~\cite{braaten2006ufb}. For $N=4$, the curve corresponds to predictions from Ref.~\cite{von2009signatures}. For $N=5$, we carry out CG calculations for the model Hamiltonian. The good agreement in the $N=3-5$ comparisons represents strong evidence of the validity of our model Hamiltonian.

The predictions from Fig.~\ref{EnNeg} imply that there is at least one $N$-body state universally related to an Efimov state.
These states are weakly bound. For example, the trimer energies are $10^5$ times smaller than the typical two-body interaction energy $\hbar^2/(m r_0^2)$.
Also, their energies follow a smooth and simple dependence on $\kappa_0 a$ that we approximate by $E_{Nb}\approx E^U_{Nb}(x+c_N\, x^{b_N})/(1+c_N)$, where $x=(a-a^*_{Nb})/a$, $E^U_{Nb}$ is the cluster energy at unitarity, and $a^*_{Nb}$ is the critical scattering length for which the cluster energy reaches the zero-energy threshold. $E^U_{Nb}$ and $a^*_{Nb}$ are extracted from the DMC results (see e.g., Table~\ref{table1}) and $c_N$ and $b_N$ are fitting parameters.
The dashed curves in Fig.~\ref{EnNeg} shows the proposed analytical expression for $E_{6b}$, $E_{7b}$, and $E_{8b}$ when setting $c_6\approx0.58$, $c_7\approx0.56$, $c_8\approx0.84$ and $b_6\approx2.68$, $b_7\approx2.86$, $b_8\approx2.50$.
The simple dependence of $E_{Nb}$ suggests that, close to unitarity, the different cluster states follow generalized Tjon lines, i.e., their energies are linearly correlated to each other and $E_{Nb}\approx \alpha_N +\beta_N\,E_{(N-1)b}$. The Tjon lines are verified with the numerics. For example, we obtain  $E_{4b}\approx 2.66 E^U_{3b} +2.0\,E_{3b}$.

The positions of the critical scattering length $a^*_{Nb}$ (dashed vertical lines in Fig.~\ref{EnNeg}) reflect the Borromean nature of these states~\cite{blume2002formation}: for $a^*_{Nb}<a<a^*_{(N-1)b}$, the $N$-body cluster is bound but all smaller clusters are unbound. The approximately equal spacing of the vertical dashed lines in  Fig.~\ref{EnNeg} suggests that the critical scattering lengths follow a simple relationship.
Actually, the critical scattering lengths for $3\le N\le8$ can be well described by $1/(\kappa_0 a^*_{Nb})\approx 2.3(1)-N$ which also suggests that $ a^*_{Nb}\rightarrow0$ when  $N\rightarrow\infty$. This prediction implies, in agreement with previous studies~\cite{bruch}, that there exist large cluster states for any $a<0$.

Next, we benchmark the energy relations at unitarity. Table~\ref{table1} presents the DMC energies of cluster states up to $N=13$ at unitarity.   To determine these energies, we analyze systems with $R_c/r_0=4$, 5, and 8 (which is equivalent to changing $\kappa_0$). The reported uncertainty is estimated by considering the dependence of the results on $R_c/r_0$ and the statistical uncertainty. Our predictions agree only qualitatively with those of Ref.~\cite{hanna2006energetics} that considers states more tightly bound and which are more affected by nonuniversal corrections.

\begin{table}
\caption{Energies at unitarity and scattering-length ratios that characterize weakly bound cluster states. The scattering length ratios can be transformed to an absolute scale using $1/(\kappa_0 a_{3b})\approx 0.64$.}
\label{table1}
\begin{center}
\begin{tabular}{||c|c|c||c|c||}
\hline
$N$ & $E^U_N/E^U_3$ & $a^*_{Nb}/a^*_{(N-1)b}$ & $N$ & $E^U_N/E^U_3$  \\ \hline
4 &  4.66(4)&  0.42(1)   & 9 & 49.9(6)      \\
5 & 10.64(4) &  0.60(1)   & 10 & 60.2(6)     \\
6 & 18.59(5) &   0.71(1)  & 11 & 70.1(7)     \\
7 & 27.9(2) &   0.78(1) & 12 & 79.9(3)    \\
8 & 38.9(3) &   0.82(1)  & 13 & 88.0(7)     \\ \hline
\end{tabular}%
\end{center}
\end{table}

The CG studies allow us to analyze the existence of excited bound states.
Four-body calculations with the model Hamiltonian predict the existence of two states whose energies at unitarity are $E^{U}_{4b}\approx 4.55E^U_{3b}$, and $E^{U,2}_{4b}\sim 1.003E^U_{3b}$ in agreement with previous studies~\cite{hammer2007upf,von2009signatures}.
 Next, we extend the study to five particles.
First, we determine the existence of a five-body state with an energy $E^{U}_{5b}\approx 10.4E^U_{3b}$, in good agreement with our QMC results. Also, we  observe an extremely weakly bound  excited five-body state. This state becomes more strongly bound as nonuniversal corrections become more important, i.e., $R_c/r_0\rightarrow0$. We conclude that there exists an second state that is bound or on the verge of being bound and that, under experimental conditions, nonuniversal corrections of the realistic interactions are going to be crucial for ultimately determining the existence of the second five-body states.
 For $N=6$, we can only analyze the ground state, but we cannot reach full convergence. However, through an analysis of the energy dependence on the basis size and an extrapolation to infinite basis functions, we estimate $E_{6b}\approx 18.3(3) E_{3b}$, in good agreement with the DMC predictions.

\begin{figure}[htbp]
\begin{center}
\includegraphics[scale=0.8,angle=0]{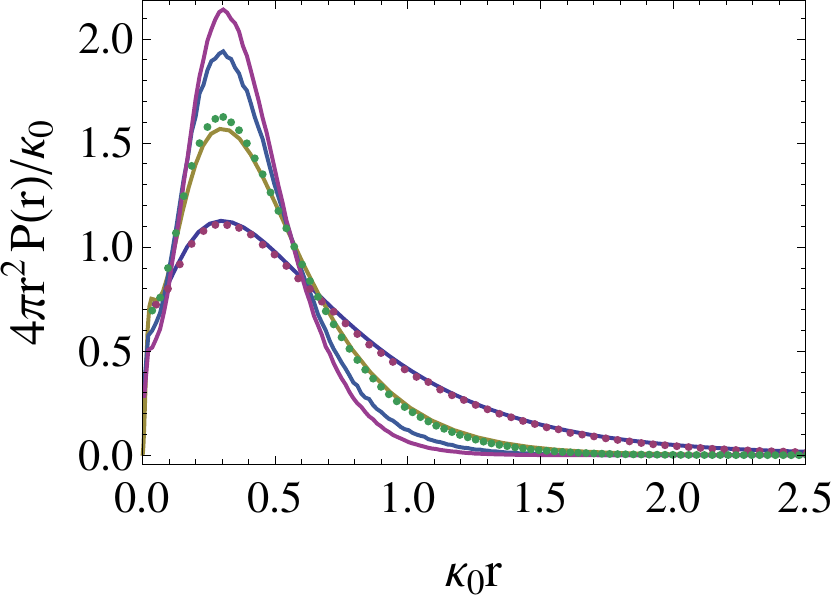}
\caption{(Color online) Pair-correlation distributions for $N=3-6$. Note that the maximum of the pair-correlation increase with $N$. For $N=3$,~4, circles correspond to the DMC predictions, while solid lines correspond to other predictions (see text). For $N=5$ and 6, the curves correspond to DMC predictions.   } \label{PCplot}
\end{center}
\end{figure}
Resonant enhancement of losses produced by the presence of a bound state at the fragmentation energy threshold has been the key signature for identifying universal three- and four-body physics~\cite{kraemer2006eeq,von2009signatures,ferlaino2009evidence}. This enhancement can also be a suitable path for experimentally observing the universal physics of larger systems. A theoretical framework for the analysis of $N$-body recombination processes has been recently proposed~\cite{mehta2009general}.
The critical scattering-length ratios reported in Table~\ref{table1} complement that study by identifying how resonant enhancements produced by different $N$-body clusters are related to each other.
The ratio $a^*_{4b}/a^*_{3b}$ is in good agreement with other theoretical predictions~\cite{von2009signatures} and experimental results~\cite{ferlaino2009evidence}. The ratio $a^*_{5b}/a^*_{4b}$ predicts that for the Cs experiment of Ref.~\cite{ferlaino2009evidence}, there should be a five-body resonant feature at about $0.6\times410\, a_0\approx 250 a_0$. The experimental results for the losses show a series of g-wave Feshbach resonances in that region that prevent us from identifying a possible five-body resonant feature.

Finally, we analyze the structural properties of these cluster states.
 Figure~\ref{PCplot} presents the pair-correlation functions, defined as $4\pi r^2P(r)=\langle\Psi|\delta(r_{ij}-r)|\Psi\rangle$, for $N=3-6$ at unitarity. In our DMC implementation, we calculate the structural properties, $P$, using the mixed estimator $\langle P\rangle_{mixed}=2\langle P\rangle_{DMC}-\langle P\rangle_{VMC}$~\cite{hamm94}.  For $N=3$, we compare the model Hamiltonian predictions calculated with DMC  (circles) with pair-correlation results extracted from semianalytical results (curve on top of circles, indistinguishable from the DMC predictions). For $N=4$, we compare the model Hamiltonian predictions calculated with DMC  (circles) with universal pair-correlation results extracted from Ref.~\cite{von2009signatures}.
 We found excellent agreement in both comparisons. The long-range binding of the states can be deduced from the pair-correlation functions which show that the  mean interparticle distance is much larger than the typical interaction  distances $r_0$ and $R_c$. For example, the peaks of the pair correlations are at $\sim7 R_c$ (i.e., $r_0\sim30 r_0$) for the parameters used in Fig.~\ref{PCplot}. 

In conclusion, we have combined a model Hamiltonian with numerical techniques to construct a suitable framework for the description of Efimov physics in many-body systems. This model description can be extended to study systems with different species and spin statistics.
 Our results present a universal picture for bosons in the $a<0$ regime and predict a series of cluster states universally related to each other.
 As a consequence of universality, the cluster states inherit the energy and size Efimov-scaling relationships that control the universal three- and four-body states. Thus, two clusters tied to two consecutive Efimov trimers would have their energies at unitarity related by $E_{Nb,n+1}=e^{-2\pi/s_0} E_{Nb,n}$.
 These predictions oppose the notion of an ``$N$-body Efimov effect''~\cite{FJ}.
 Our predictions can be experimentally investigated by analyzing resonant losses in ultracold bosonic gases.
Alternatively, these cluster states can be created by loading bosonic atoms in an optical lattice. The proposal of Ref.~\cite{stoll2005production} of creating trimers in optical lattices can be extended to larger systems. In principle, the energies of different cluster states can be measured as functions of $a$, experimentally reproducing the results of Fig.~\ref{EnNeg}.

Enlightening discussions with C. H. Greene, D. Blume, and J. P. D'Incao are acknowledged. This work was supported by NSF.

\end{document}